\documentclass{article}
\usepackage{spconf,amsmath,graphicx}
\usepackage{xcolor}
\usepackage{amsmath} 
\usepackage{amssymb}
\usepackage{hyperref}
\usepackage{booktabs}
\usepackage{enumitem}
\setlist{nosep, leftmargin=14pt}

\usepackage{mwe} 

\title{RibCageImp: A Deep Learning Framework for 3D Ribcage Implant Generation}
\name{Gyanendra Chaubey$^{1}$, Aiman Farooq$^{2}$, Azad Singh$^{2}$, and Deepak Mishra$^{1,2}$}
\address{
    $^{1}$ School of AI and Data Science, IIT Jodhpur, RJ 342030 India \\
   $^{2}$ Department of Computer Science and Engineering, IIT Jodhpur, RJ 342030, India\\
   Email:- {\{m23air005, farooq.1, singh.63, dmishra\}} @{\{iij.ac.in\}}
}
\begin{document}
%
\maketitle
\begin{abstract}
The recovery of damaged or resected ribcage structures requires precise, custom-designed implants to restore the integrity and functionality of the thoracic cavity. Traditional implant design methods rely mainly on manual processes, making them time-consuming and susceptible to variability. In this work, we explore the feasibility of automated ribcage implant generation using deep learning. We present a framework based on 3D U-Net architecture that processes CT scans to generate patient-specific implant designs. To the best of our knowledge, this is the first investigation into automated thoracic implant generation using deep learning approaches. Our preliminary results, while moderate, highlight both the potential and the significant challenges in this complex domain. These findings establish a foundation for future research in automated ribcage reconstruction and identify key technical challenges that need to be addressed for practical implementation.

\end{abstract}
\section{Introduction}
\label{sec:intro}

The reconstruction of compromised thoracic structures, such as ribcage, requires high-precision engineering in implant design. Accurate dimensional specifications are crucial for restoring anatomical alignment and biomechanical functional integrity of the chest cavity \cite{moradiellos2017functional}. While modern medical imaging and advanced technologies have enabled the development of many prosthetic structures, ribcage reconstruction presents unique challenges \cite{ciraulo1994flail}. The complex 3D geometry of the thoracic cavity and patient-specific anatomical variations make the design of custom ribcage implants particularly complex. The conventional design workflows remain primarily manual and require extensive specialized expertise \cite{artec_chest_implants}. It requires extensive time for clinicians to analyze patient-specific thoracic architecture through manual measurements and Computer-Aided Design (CAD) interpretations, translating complex anatomical data into fabrication-ready implant specifications. These technical limitations and the critical need for precise anatomical alignment create significant barriers to delivering optimal patient-specific thoracic implants. 



The primary challenge lies in achieving high accuracy and efficiency in ribcage implant design while minimizing manual interventions. This necessitates the development of automated design approaches that can maintain thoracic geometric precision, reduce processing time, and ensure consistent quality across different patients. Additionally, such automation must account for critical anatomical landmarks, biomechanical constraints, and surgical considerations that traditionally rely on clinical expertise.
\begin{figure}[htbp]
    \centering
    \includegraphics[width=\columnwidth]{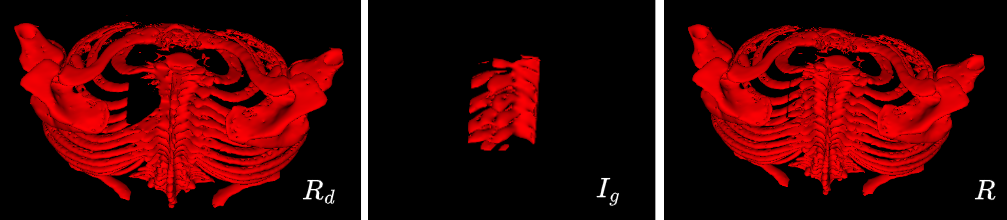}
    \caption{Illustration of defective ribcage $R_d$, ground truth implant $I_g$, and complete ribcage with predicted ground truth $R_d+I_g$.}
    \label{fig:illustration}
\end{figure}

The emergence of deep learning methods such as Convolutional Neural Networks (3D CNNs) \cite{tran2015learning}, especially the 3D U-Net~\cite{cciccek20163d}, has demonstrated remarkable success in medical image analysis. These networks have shown capabilities in processing volumetric medical data for complex tasks, including brain tumor segmentation \cite{mehta20183d}, early disease detection and diagnosis \cite{liu2021development}, and surgical planning \cite{wang2024nnu}. While 3D U-Net-based models have been successfully implemented for cranial implant design \cite{wodzinski2022deep}, the ribcage presents significantly more complex challenges. Cranial implants primarily deal with relatively uniform, curved surfaces focused mainly on structural protection and aesthetic outcomes.  In contrast, ribcage implants must address multiple curved bones with varying cross-sections, intricate joint connections, and complex spatial relationships. Further, it also ensures the functionality of proper respiratory mechanics, vital organ protection, natural chest wall movement, and seamless integration with the surrounding musculoskeletal structure.

Ribcage reconstruction becomes important in various medical situations. These include the removal of chest tumors (both primary and spreading cancers), severe chest injuries from accidents, and congenital disabilities affecting chest wall structure. Each patient's case is unique and requires a customized implant solution that perfectly matches their specific anatomy \cite{icsik2021reconstruction}. Developing automated, learning-based approaches for ribcage implant design promises significant benefits. This advancement reduces design time, improves implant quality, and enhances surgical outcomes and patient recovery. This is particularly significant given the time-critical nature of many cases, especially in emergency trauma scenarios requiring rapid implant design and efficient surgical planning in oncological cases. An extensive literature review reveals a complete absence of learning-based approaches for automated ribcage implant design, with current methods relying solely on manual CAD processes, statistical shape analysis, template-based modifications, and expert-guided design iterations \cite{JOVS32520}.

CT scans offer high-resolution, volumetric views of thoracic anatomy, ideal for developing learning-based methods for direct 3D implant generation. Leveraging this rich anatomical information—such as spatial relationships and bone structure—into a 3D U-Net model for implant design could seamlessly integrate into standard clinical workflows. 
Our key contributions to the paper are:
 
\begin{itemize}
    \item We propose the first deep-learning approach for automated ribcage implant generation directly from CT scans. This establishes a new paradigm for thoracic reconstruction, moving beyond traditional CAD-based methods towards automated, data-driven solutions. 
    \item We adapt and optimize the 3D U-Net architecture to demonstrate the feasibility of ribcage reconstruction. 
    \item We provide comprehensive evaluation and analysis through detailed validation of the proposed approach on diverse patient cases. The current results demonstrate the viability of the deep model and highlight opportunities for further refinement and optimization.
\end{itemize}
This work provides a preliminary solution and opens new research directions for advancing automated ribcage implant generation. It serves as a foundational reference for future researchers aiming to contribute to this domain.

    
    


\section{Methodology}
\label{sec:method}


Given a volumetric CT scan $S \in \mathbb{R}^{W \times H \times D}$, where $W$, $H$, and $D$ represent the width, height, and depth dimensions of the scan, we aim to generate a 3D implant for a defective region in the ribcage. Let $R$ represent an intact ribcage structure. When a portion requires removal (either due to surgical access to underlying organs or due to fracture), the remaining structure is represented as $R_d \subseteq S$, and the portion requiring reconstruction serves as the ground truth implant $I_g$. The aim is to generate a 3D implant $I_p \in \mathbb{R}^{W' \times H' \times D'}$, where $W' < W$, $H' < H$, and $D' < D$, that when combined with $R_d$, reconstructs the complete ribcage $\hat{R} = R_d + I_p$. The challenge of finding the optimal $I_p$ can be formalized as:


\begin{equation}
    \min_{R_d} \; \mathcal{L}(I_g, I_p) \quad \text{subject to} \quad \hat{R} = R_d + I_p
    \label{eq:ribcage2}
\end{equation}
\noindent Where $\mathcal{L}$ is a loss function that evaluates the similarity between the predicted and ground truth ribcage implants. 



\begin{figure}[h]
    \centering
    \includegraphics[width=\columnwidth]{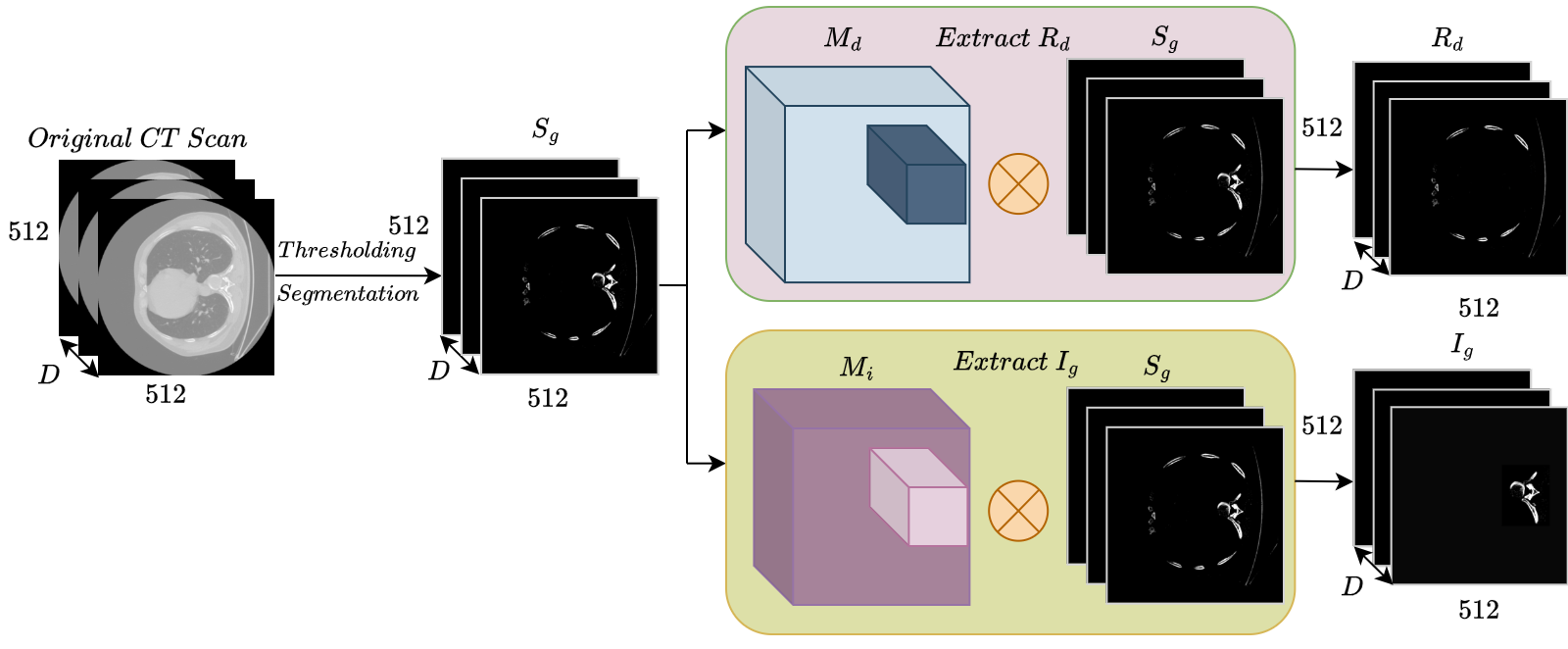}
    \caption{Data preparation for ribcage reconstruction, showing thresholding and segmentation of CT scans to obtain the defective region \( R_d \) and the implant \( I_g \) for model training.}
    \label{fig:data_preparation}
    \vspace{-0.5cm}
\end{figure}

\begin{figure*}[htbp]
    \centering
    \includegraphics[width=\textwidth]{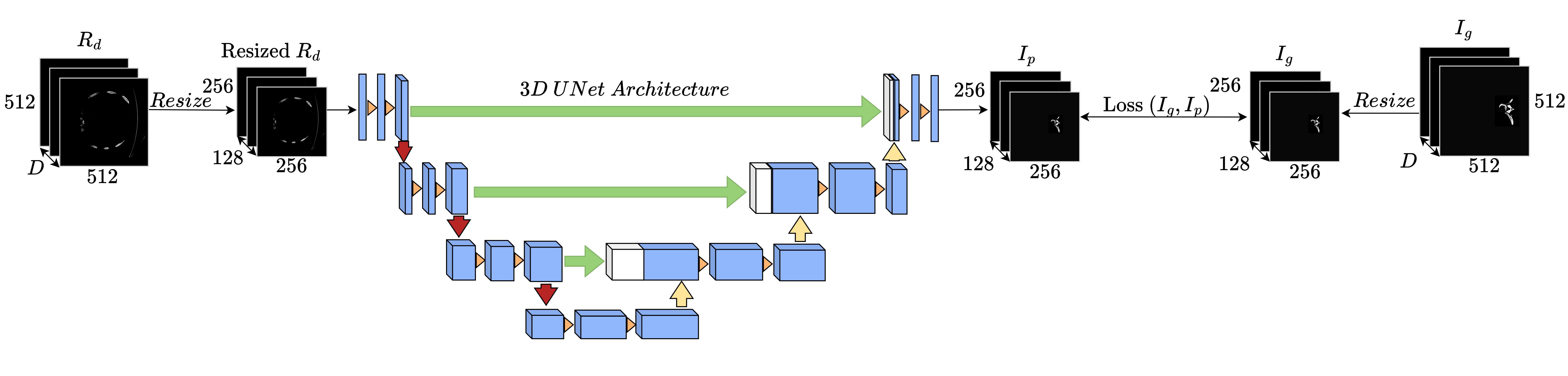}
    \caption{Training pipeline for ribcage implant generation using a 3D U-Net. The defective region \( R_d \) and ground truth implant \( I_g \) are processed and input into the network, which outputs the predicted implant \( I_p \). The prediction is evaluated against \( I_g \) with a loss function \( L(I_g, I_p) \) for reconstruction accuracy.}
    \label{fig:training}
\end{figure*}

\subsection{Data Preparation}
\label{ssec:datapreprocess}
In this work, we use the RibFrac dataset from the RibFrac Grand Challenge 2020~\cite{ribfracclinical2020}, consisting of 420 CT samples for training, 160 for testing, and 80 for validation. To isolate bones, we apply value-based thresholding on the CT scans, resulting in segmented bones denoted as \( S_g \). A corresponding mask of the CT volume’s dimensions is generated to retain relevant pixel values. For implant generation, we create a defect mask \( M_d \) by embedding a zero mask of size \( 64 \times 64 \times 64 \) in selected regions between heights 0.5 and 0.75, ensuring sternum inclusion. The implant mask \( M_i \) is derived by subtracting \( M_d \) from 1 and multiplying by \( S_g \), resulting in the ground truth implant \( I_g \). Figure~\ref{fig:illustration} shows the defective ribcage, complete ribcage, and implant region, while Figure~\ref{fig:data_preparation} outlines the data preparation workflow. Extensive data preparation techniques further enhance the model's performance while thresholding and masking isolate bone structures and simulate defect regions in a standardized way.


\subsection{Network Architecture}
\label{ssec:proposedsolution}

The training architecture is illustrated in Figure~\ref{fig:training}. The data is loaded with normalization and resizing to the fixed size of 256 $\times$ 256 $\times$ 128 to maintain data consistency for the network. 
The proposed solution leverages a deep learning framework based on the 3D U-Net architecture and an EfficientNet-b0 encoder for automated ribcage implant generation. 3D U-Net's architecture is particularly well suited for processing volumetric CT data, enabling effective feature extraction and reconstruction of damaged thoracic regions.  
The framework employs Mean Squared Error (MSE) Loss for voxel-wise optimization, ensuring precise alignment with existing structures while preserving patient-specific anatomical features.
This automated approach significantly reduces the variability inherent in traditional manual and CAD-based methods while maintaining anatomical fidelity in the generated implants.

\subsection{Loss functions}
To ensure anatomically accurate implant generation, we employ a combination of complementary loss functions:


 
\noindent \textbf{Dice Loss \cite{milletari2016vnet}} to measure the volumetric overlap between the predicted implant $I_p$ and the ground truth implant $I_g$ as \ref{eq:ribcage4}.

\begin{equation}
\mathcal{L}_{\text{Dice}}(I_p, I_g) = 1 - \frac{2 \cdot |I_p \cap I_g|}{|I_p| + |I_g|}
    \label{eq:ribcage4}
\end{equation}
   
\noindent \textbf{MSE Loss}, to quantifies the voxel-wise accuracy of the prediction \ref{eq:ribcage5}.
\begin{equation}
 \mathcal{L}_{\text{MSE}}(I_p, I_g) = \frac{1}{n} \sum_{i=1}^{n} \left( I_p(i) - I_g(i) \right)^2
    \label{eq:ribcage5}
\end{equation}
\noindent where $n$ represents the number of voxels in the implant region. MSE loss ensures precise voxel-level alignment.
\\
\noindent \textbf{Extra Region Removal (ERR) loss,} that penalizes any extraneous predictions \(E_{\text{R}}\) outside the ground truth region (as in Equation \ref{eq:ribcage6}) in the predicted implant \( I_p \) by computing the elementwise product with the inverse of the ground truth \( I_g^{\text{-1}} \) and subtracting to a zero matrix, as in Equation \ref{eq:ribcage7}.
\begin{equation}
E_{\text{R}} = \left( I_g^{-1} \circ I_p \right)
\label{eq:ribcage6}
\end{equation}

\begin{equation}
\mathcal{L}_{\text{ERR}}(E_{\text{R}}, \mathbf{0}) = \frac{1}{n} \sum_{i=1}^{n} \left( E_{\text{R}}(i) - \mathbf{0}(i) \right)^2
\label{eq:ribcage7}
\end{equation}
where \(\mathbf{0}\) is a zero-valued matrix with same dimensions as \( I_p \).

\noindent \textbf{Gap Filling (GF) loss,} which addresses structural discontinuities, penalizing missing regions in the prediction. The gaps $G_{\text{F}}$ are identified as: 
\begin{equation}
G_{\text{F}} = I_p^{-1} \circ I_g
\label{eq:ribcage8}
\end{equation}

\noindent The GF loss is computed as:
\begin{equation}
\mathcal{L}_{\text{GF}}(G_{\text{F}}, \mathbf{0}) = \frac{1}{n} \sum_{i=1}^{n} \left( G_{\text{F}}(i) - \mathbf{0}(i) \right)^2
\label{eq:ribcage9}
\end{equation}

\noindent We subtract from the zero matrix in both ERR and GF for smooth gradient flow, effective penalization on loss, and maintaining consistency with the MSE loss. Each loss component addresses a specific aspect of implant accuracy: Dice loss for overall shape, MSE for voxel precision, ERR for preventing overextension, and GF for ensuring completeness.
The complete optimization objective combines all loss components: $\mathcal{L}_{\text{rib}} =  \mathcal{L}_{\text{MSE}} + \mathcal{L}_{\text{ERR}} + \mathcal{L}_{\text{GF}}$

\section{Experimental Results \& Discussions}
\label{sec:experiments}
We evaluated our proposed approach for automated ribcage implant generation using a 3D U-Net architecture with an EfficientNet-B0 encoder pre-trained on ImageNet. All the experiments are performed on the RibFrac dataset comprising 300 training samples and 160 test cases to ensure robust validation of the model's performance. For quantitative evaluation, we employ two complementary metrics: Dice score coefficient (DSC) \cite{milletari2016vnet} to measure volumetric overlap between predicted and ground truth implants, and Hausdorff Distance (HD)~\cite{huttenlocher1993comparing} to assess surface accuracy and geometric similarity. The network is trained using the Adam optimizer with a learning rate of 1e-5 and weight decay of 1e-4, utilizing a batch size of 2. All experiments are implemented in PyTorch and conducted on an NVIDIA RTX A5000 GPU.

 \begin{table}[!t]
    \centering
    \begin{tabular}{l|c|c}
        \toprule
        \textbf{Model with Loss Function} & \textbf{DSC} & \textbf{HD(mm)} \\
        \midrule
        U-Net with DICE & 0.0814 & 226.18 \\
        U-Net with MSE & 0.1615 & 220.94 \\
        U-Net with MSE + ERR & 0.2233 & 206.67 \\
        \bf{U-Net with MSE + ERR + GF} & \bf{0.2524} & \bf{148.90} \\
        \bottomrule
    \end{tabular}
    \caption{Quantitative test set evaluation of ribcage implant generation using DSC and HD metrics across different configurations of the loss functions.}
    \label{tab:results}
\end{table}




\subsection{Quantitative Results}
The experimental evaluation incorporates multiple loss function configurations: Dice, MSE, MSE with ERR, and MSE combined with both ERR and GF. Table~\ref{tab:results} presents quantitative results across different loss configurations, comparing their performance using DSC and HD metrics. The model demonstrates promising results with MSE configurations, achieving a DSC of \textbf{0.52} in specific implant generation scenarios. These initial results, while moderate, highlight both the potential of automated implant generation. The moderate results can be attributed to several key challenges: the complex three-dimensional geometry of the ribcage with varying curvatures and intersections and the significant anatomical variations across patients Additionally, the task of implant generation requires not only structural reconstruction but also the preservation of subtle anatomical features crucial for maintaining thoracic functionality. Despite these challenges, the results demonstrate the feasibility of deep learning-based approaches for automated ribcage implant generation, establishing a foundation for future improvements.

\subsection{Qualitative Visualizations}
\begin{figure}[!t]
    \centering
    \includegraphics[width=\columnwidth]{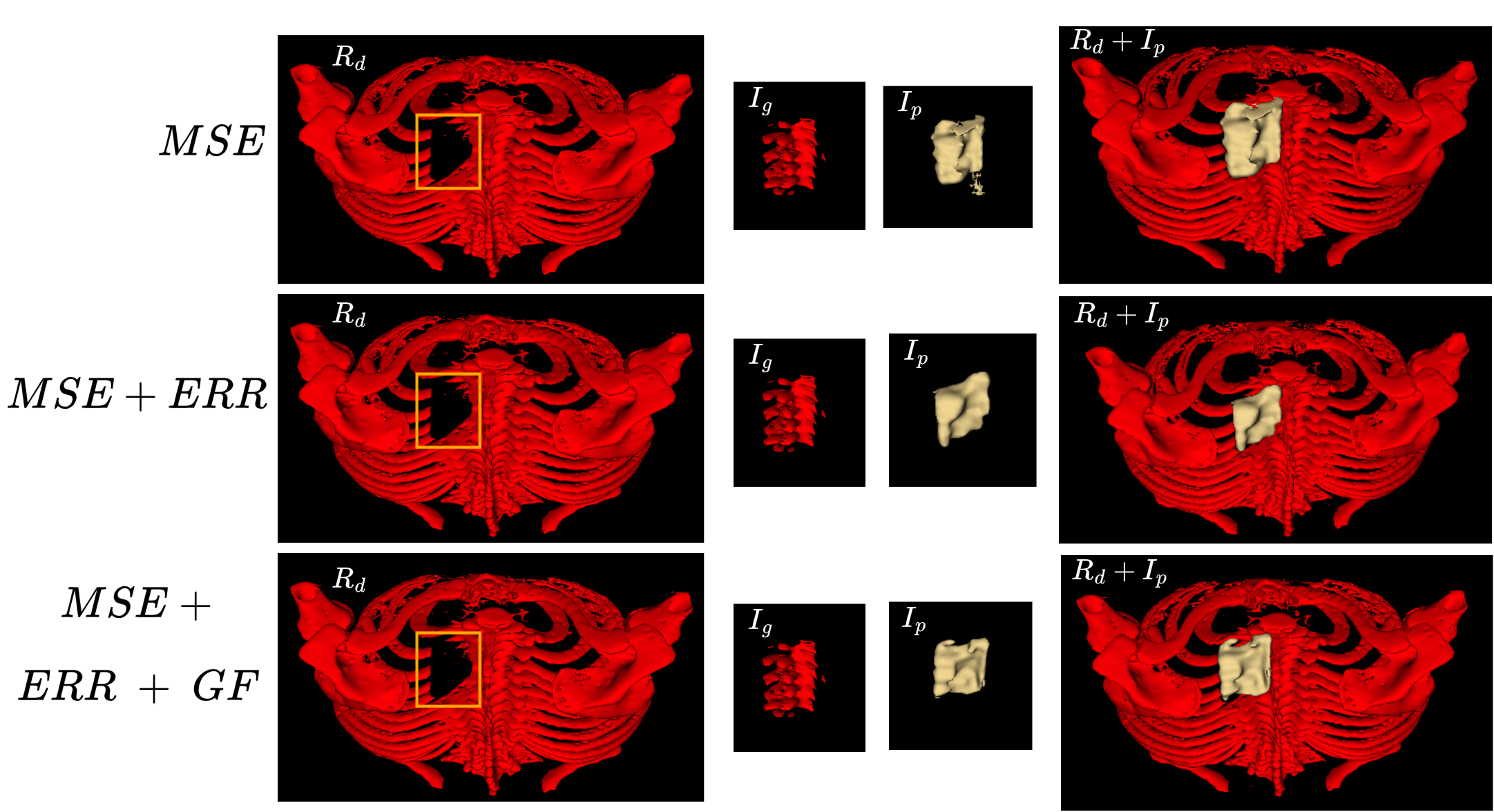}
    \caption{Implant prediction results on a defective ribcage, showing the ground truth and the fitted ribcage with the predicted implant using different configurations of loss terms.
    }
    \label{fig:results}
\end{figure}

Figure~\ref{fig:results} presents a comparative visualization of the ground truth and predicted implant structures within a complex thoracic region. While the quantitative metrics indicate room for improvement, the visual comparison provides insights into the model's current reconstruction behavior and its limitations. The predicted structures show initial promise in capturing basic anatomical features, though achieving precise alignment with the intricate geometry of the original bone structure remains challenging. 

\begin{figure}[ht]
    \centering
    \includegraphics[width=\columnwidth]{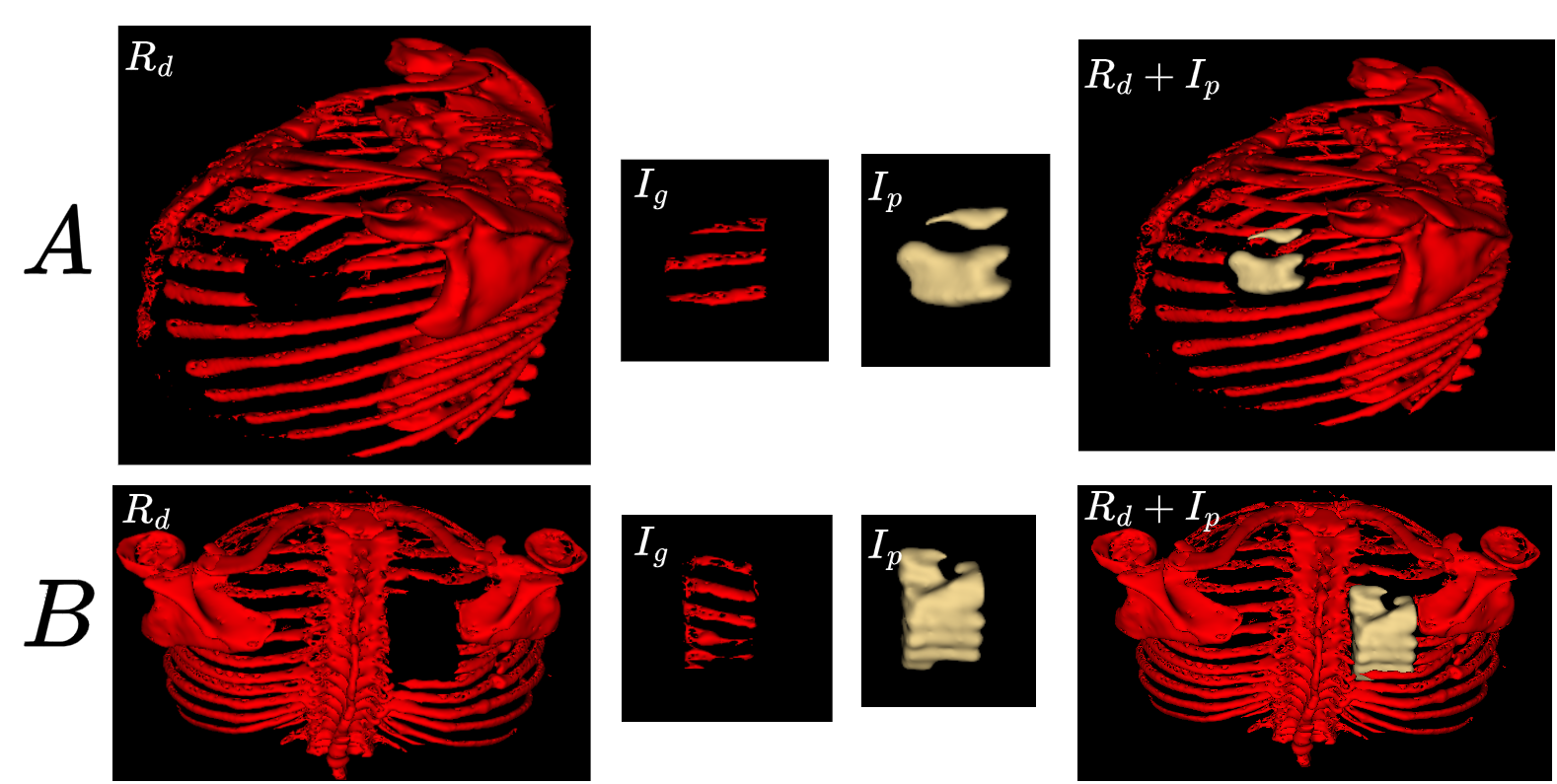}
    \caption{Comparison of ground truth and predicted implant regions showing typical failure cases (A and B) with extraneous regions and structural gaps, highlighting areas for improvement in network design and loss functions.
}
   \vspace{-0.5cm}
    \label{fig:wrongpred}
\end{figure}

To better understand these limitations, Figure~\ref{fig:wrongpred} illustrates two specific cases (A and B) using our best-performing configuration (MSE+GF+ERR). The predictions exhibit two key challenges: extraneous regions where the predicted implant extends beyond the intended boundaries and structural gaps where the implant fails to fully cover the defect. These examples highlight that ribcage implant prediction is particularly challenging due to the complex nature of thoracic anatomy. The ribcage's intricate structure, featuring curved surfaces, varying bone thicknesses, and intricate intersections between ribs and surrounding structures, makes it significantly more complex than simpler anatomical reconstructions like cranial implants. The challenge lies not only in capturing the overall structural geometry but also in maintaining precise adherence to local anatomical constraints, including proper curvature, thickness variations, and connections with adjacent ribs. These preliminary visual results, combined with our quantitative metrics, emphasize the need for advanced approaches that can better handle the multi-scale geometric complexity of thoracic structures, from global ribcage architecture to fine-grained local features essential for functional reconstruction.


\section{Conclusion}
\label{sec:discussions}

This study establishes the feasibility of automated thoracic reconstruction and identifies key challenges that need to be addressed in this domain. We presented a preliminary solution for automated ribcage implant generation to address the limitations of manual and CAD-based methods. Using a 3D U-Net architecture to process CT scan data, we explored the feasibility of deep learning for patient-specific implant generation. Our initial results demonstrate both the complexity of the problem and the challenges in capturing precise anatomical details. The moderate performance metrics reflect the inherent difficulty in accommodating patient-specific variations and complex thoracic geometry. Future work focuses on expanding the training dataset and exploring advanced architectures and loss functions for improved geometric accuracy.



\bibliographystyle{IEEEbib}
\bibliography{strings}

\end{document}